\documentclass[superscriptaddress,reprint, aps, prx,twocolumn]{revtex4-2}
\usepackage[T1]{fontenc}
\usepackage{times}
\usepackage{pslatex}
\usepackage{graphicx}
\usepackage{float}
\usepackage[sort&compress]{natbib}
\usepackage{amsmath,scalerel}
\usepackage[colorlinks=true,linkcolor=blue, citecolor=blue]{hyperref}%
\usepackage{natbib}
\usepackage{tabularx}
\usepackage{graphicx}
\usepackage{bm}
\usepackage{color}
\usepackage{enumitem}
\usepackage{amssymb}

\begin{document}
\title{Sensing of magnetic field effects in radical-pair reactions using a quantum sensor}
\author{Deepak Khurana}
\email{deekh@dtu.dk}
\author{Rasmus H. Jensen}
\author{Rakshyakar Giri}
\author{Juanita Bocquel}
\author{Ulrik L. Andersen}
\affiliation{Center for Macroscopic Quantum States (bigQ), Department of Physics, Technical University of Denmark, 2800 Kgs. Lyngby, Denmark}
\author{Kirstine Berg-S{\o}rensen}
\affiliation{Department of Health Technology, Technical University of Denmark, 2800 Kgs. Lyngby, Denmark}{}
\author{Alexander Huck}
\email[]{alhu@dtu.dk}
\affiliation{Center for Macroscopic Quantum States (bigQ), Department of Physics, Technical University of Denmark, 2800 Kgs. Lyngby, Denmark}

\begin{abstract}
Magnetic field effects (MFE) in certain chemical reactions have been well established in the last five decades and are attributed to the evolution of transient radical-pairs whose spin dynamics are determined by local and external magnetic fields. The majority of existing experimental techniques used to probe these reactions only provide ensemble averaged reaction parameters and spin chemistry, hindering the observation of the potential presence of quantum coherent phenomena at the single molecule scale. Here, considering a single nitrogen vacancy (NV) centre as quantum sensor, we investigate the prospects and requirements for detection of MFEs on the spin dynamics of radical-pairs at the scale of single and small ensemble of molecules. We employ elaborate and realistic models of radical-pairs, considering its coupling to the local spin environment and the sensor. For two model systems, we derive signals of MFE detectable even in the weak coupling regime between radical-pair and NV quantum sensor, and observe that the dynamics of certain populations, as well as coherence elements, of the density matrix of the radical pair are directly detectable. Our investigations will provide important guidelines for potential detection of spin chemistry of bio-molecules at the single molecule scale, required to witness the hypothesised importance of quantum coherence in biological processes.
\end{abstract}

\maketitle

\section{Introduction}
Investigating the fundamental role of spin interactions in magnetic field effects (MFE) in chemical reactions has a long history~\cite{steiner1989magnetic, woodward2002radical, timmel2004study}. The study of spin-chemical effects provides important insights about structure, kinetics and magnetic properties of transient intermediate chemical species~\cite{hayashi2004introduction, hore2020spin} and are explored in various interdisciplinary applications including sensitivity enhancement in nuclear magnetic resonance (NMR)~\cite{lee2014sensitivity}, quantum computing~\cite{rugg2019photodriven}, avian magnetic compass \cite{hore2016radical}, and solar energy conversion kinetics in photosynthetic systems~\cite{colvin2011magnetic}. 

It is astounding that spin interactions can have decisive effects on the fate of chemical reactions as the energy of spin transitions typically is orders of magnitude smaller than the thermal energy~\cite{hayashi2004introduction}. However, spin dependent MFEs in chemical systems, also known as the radical-pair mechanism (RPM)~\cite{woodward2002radical, hore2016radical, rodgers2009chemical}, rely on the creation of transient paramagnetic species in a non-equilibrium state called radicals, which are chemical species with an odd number of electrons. In the RPM, the radical pair (RP) is spatially separated but in a spin-correlated state and the recombination of the radicals back to the molecular precursor state is spin selective. The influence of an external magnetic field then occurs in terms of a modulation of the spin dynamics and consequently an alteration in the yield of products formed from the various spin states~\cite{evans2013magnetic,jones2016magnetic}.

The formulation of the RPM started in the late 1960s to explain non-equilibrium  magnetic resonance spectra of chemical reactions of organic molecules~\cite{kaptein1969chemically1,kaptein1969chemically, ward1967nuclear, fessenden1963electron, closs1969mechanism, closs1970theory, wong1973chemically}, while the recent interest is fuelled by investigations of the RPM as the most plausible mechanism for MFE in biological reaction kinetics \cite{kim2021quantum, marais2018future}. These include flavoproteins related to DNA photolyases, the involvement of cryptochromes in circadian rhythms \cite{emery2000unique, panda2002circadian} and their proposed role in animal magnetoreception~\cite{wiltschko2021magnetic}. MFEs have been recorded in cryptochromes \cite{evans2013magnetic} and seem to fulfil the structural and dynamical requirements of the RPM~\cite{hore2016radical}. Considerable interest exists also in the role of radical pairs in chemical kinetics in photosynthetic reaction centres~\cite{colvin2011magnetic, marais2015quantum, kominis2013quantum}. Flavoenzymes \cite{joosten2007flavoenzymes} - Flavin-based enzymes - are responsible for catalytic functions in diverse biological reactions \cite{pimviriyakul2020overview} and the involvement of various RPs in these reactions is debated in recent discussions \cite{messiha2015magnetic, crotty2012reexamination}.
\begin{figure}
	\begin{center}
		\includegraphics[trim=9.4cm 9.3cm 14.3cm 2.5cm,clip,width=6.5cm]{{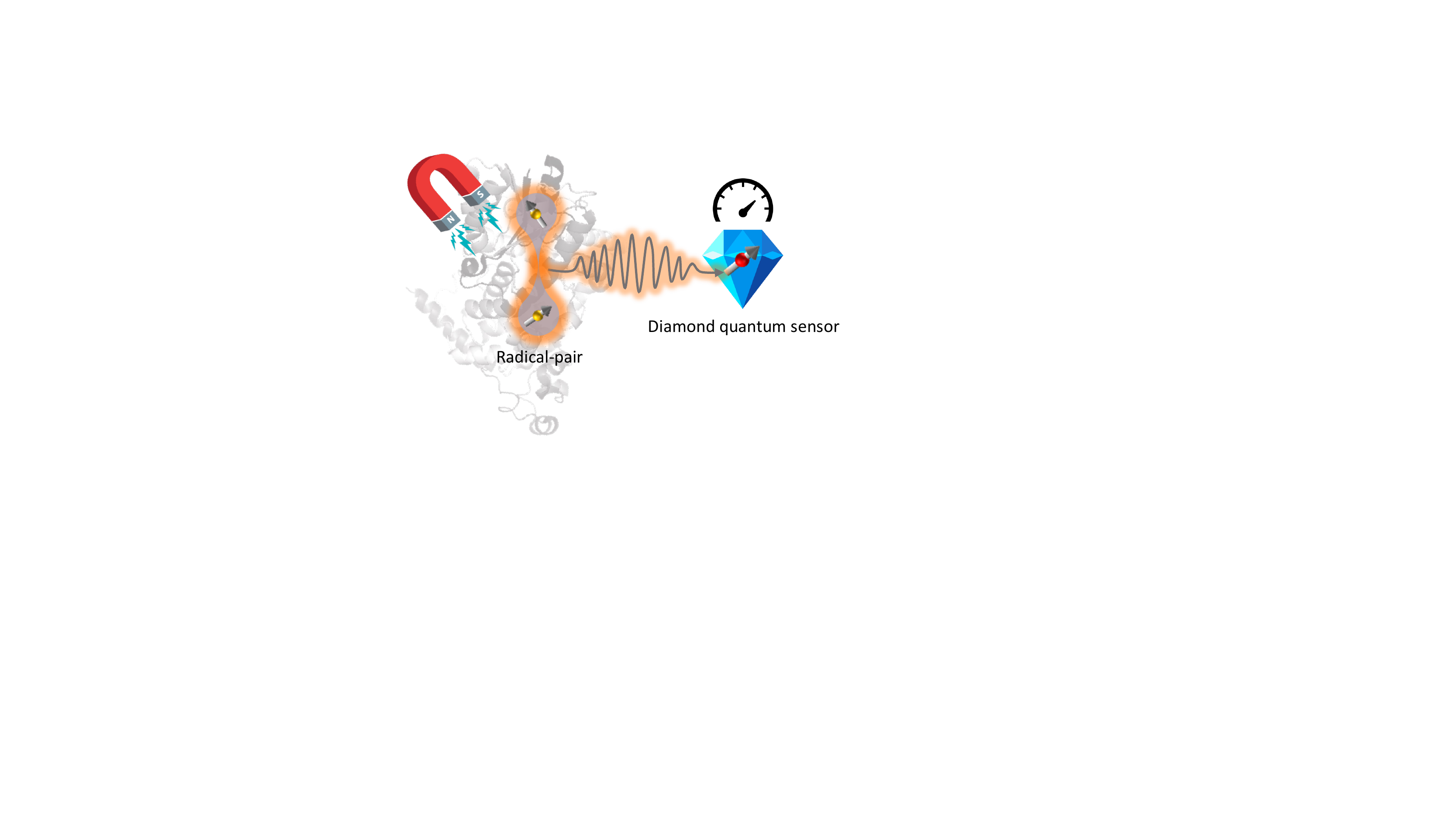}}
	\end{center}
	\caption{\small Simplified illustration for sensing of MFEs of a RP reaction using a NV quantum sensor in diamond. A RP on a host molecule is formed in a spin-correlated state (yellow). Subject to local and external fields, the distinct RP evolution induces a signal that is detectable by a nearby NV centre (red) using appropriate sensing sequences.}
	\label{schematics}
\end{figure}

Existing experimental techniques for \textit{in vitro} probing of MFE in RP dynamics in chemical reactions rely on averaged signals collected from a large ensemble of molecules. These techniques include time-resolved electron paramagnetic resonance (TREPR) spectroscopy \cite{bittl2005transient, biskup2013time}, and optical methods based on absorption \cite{henbest2008magnetic, maeda2011following, maeda2012magnetically, sheppard2017millitesla} and fluorescence detection \cite{dodson2015fluorescence, evans2015sensitive, kattnig2016chemical, dejean2020detection, ikeya2021cellular}. Studies of RP reactions conducted using these techniques can only provide ensemble averaged information of spin dynamics along with requiring a large quantity (few microliters) of potentially precious biological samples. It is therefore imperative to instead consider single-molecule detection techniques in order to reveal potential quantum coherent spin evolution otherwise hidden by ensemble averaging~\cite{ikeya2019single, liu2017scheme, finkler2021quantum}. Specifically the single negatively charged nitrogen-vacancy (NV) center in diamond~\cite{doherty2013nitrogen} has attracted significant interest as potential candidate for the detection of spin-chemical effects of RP reactions at the single molecule scale~\cite{liu2017scheme, finkler2021quantum}. This is achieved due to the excellent bio-compatibility of diamond~\cite{schirhagl2014nitrogen, simpson2019quantum}, the attainable nanometer scale spatial resolution and remarkable sensitivity to external electromagnetic fields of a single NV center~\cite{balasubramanian2008nanoscale, maze2008nanoscale, taylor2008high, dolde2011electric, dolde2014nanoscale}. 

In the present work, we discuss a realistic avenue for the detection of MFE associated with RPs at the scale of single and small ensembles ($\leq 100$) of molecules using a single NV center in diamond (Fig.\ref{schematics}). We investigate both the weak and the strong coupling regime between the NV center and RPs. Compared to the simplified model employed by Finkler et al.~\cite{finkler2021quantum} neglecting important spin interactions and considering only the strong coupling regime, we here use an elaborate model governing realistic RP spin dynamics. We include up to three nuclear spins per radical with maximum anisotropic hyperfine coupling along with dipolar and exchange coupling between the radicals. Applying standard sensing protocols, we derive measurable signals received by an NV centre and quantify magnetic field dependent RP dynamics. We show that in the weak coupling regime, the signal generated by an RP on a single-molecule is comfortably within the sensitivity limit of state-of-the-art shallow NV centers and dynamics  of certain populations, as well as coherences, of the RP density matrix is directly accessible. Non-trivial features of RP spin dynamics can be observed depending on direction and magnitude of a bias magnetic field. Further, we observe that signal features tend to average out when we consider a small ensemble of RPs, highlighting the importance of single-molecule detection. In the strong coupling regime, we find that although there is an opportunity to probe various RP spin state populations individually, detection becomes challenging especially in larger bio-molecules due to the increased number of unavoidable and spectrally indistinguishable hyperfine interactions within the RP. 

The article is organized as follows. In section~\ref{model}, we introduce the theoretical framework for the RP and interaction with the NV center. In section~\ref{sim}, we describe the numerical simulations performed with realistic RP systems and present the main results. We conclude in section~\ref{conclude} and discuss possible future directions.

\section{Description of the model}
\label{model}
\begin{figure}
	\begin{center}
		\includegraphics[trim=9.4cm 1.2cm 10.5cm 7.2cm,clip,width=8cm]{{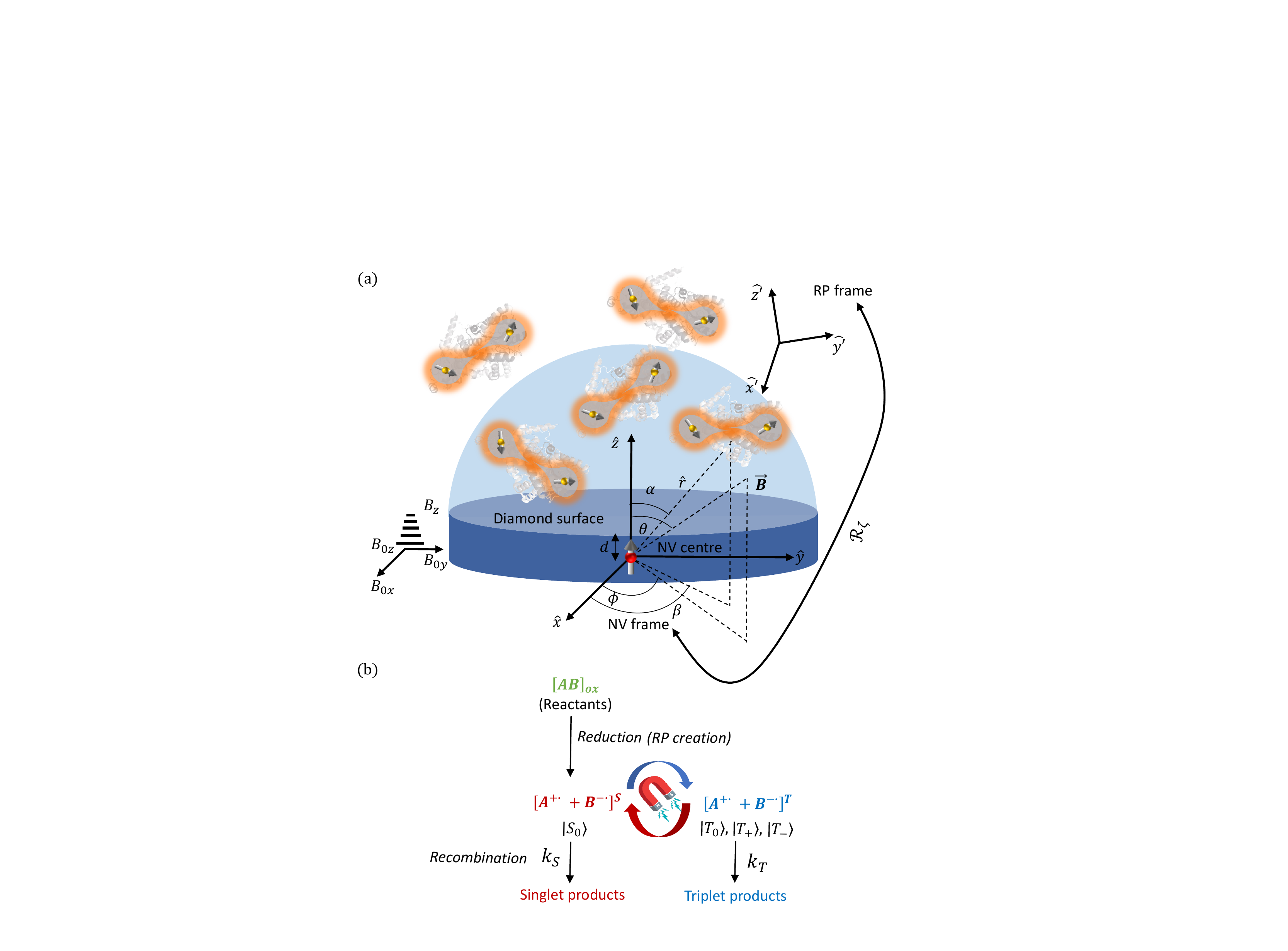}}
	\end{center}
	\caption{\small (a) General setting for detection of RP spin dynamics using an NV center situated at depth $d$ under diamond surface. The light blue hemisphere shows the sensing volume of the NV center and a RP situated in this volume contribute to the detectable signal by the NV center. Without loss of generality, we choose to work in the axis system where normal to the diamond surface is aligned along the NV axis ([1 1 1] crystal axis), named `NV frame' here. The center of the coordinate system associated with the molecule that hosts a RP is situated at location ( $r, \zeta = \{\alpha, \beta\}$) with respect to the NV center, named `RP frame'. The two axis systems are related by the rotation matrix $\mathcal{R}_\zeta$ which is different for each RP. The magnetic field is applied in direction ($\theta, \phi$). When RP spin dynamics is probed at very low magnetic fields (close to Earth's), a gradient is required for the magnetic field in the $z$ direction to ensure the two level approximation of NV center energy levels is valid. (b) Simplified description of the RPM. The two factors, (i) magnetic field dependent inter-conversion between the singlet state and triplet states, and (ii) spin dependent product yield, make the RPM a plausible mechanism behind MFE in RP reactions.  }
	\label{setting}
\end{figure}
The general situation considered for the model is illustrated in Fig. \ref {setting} (a). Molecules hosting RPs are distributed above the diamond surface  and an NV center is situated at a depth $d$ from the surface. We choose to work in the NV frame (shown in the figure) and the total Hamiltonian of the whole system in this frame is given as
\begin{align}
	H = H_{\mathrm{NV}} + H_{\mathrm{RP}} + H_{\mathrm{C}},
\end{align} 
where $H_{\mathrm{NV}}$, $H_{\mathrm{RP}}$, and $H_{\mathrm{C}}$ are the Hamiltonians governing the dynamics of the NV center, the RP, and coupling between the NV center and the RP, respectively.

We consider the $C_{3v}$ symmetric, negatively charged NV$^-$ center (hereafter called NV) that consists of ground ($^3$A, total spin $s = 1$), metastable ($^1$A, $s = 0$) and excited ($^3$E, $s = 1$) states \cite{doherty2013nitrogen}. The ground state triplet $\{|0\rangle, |\pm 1\rangle|\}$ is split at zero magnetic field with $D_{\mathrm{ZFS}} = 2.87$ GHz \cite{balasubramanian2008nanoscale}. The $|+1\rangle$ and $|-1\rangle$ can be further split by application of an external magnetic field, thus making all the three states accessible by application of appropriate control fields. Upon optical excitation typically with 532nm light, the NV center might relax via a spin-dependent inter-system crossing \cite{goldman2015state}, a process that allows for initialization in $|0\rangle$ and spin-state dependent fluorescence contrast between $|0\rangle$ and the $|\pm 1\rangle$ states. 
These properties combined with an external magnetic field can be exploited to isolate $|0\rangle \leftrightarrow |+1\rangle$ or $|0\rangle \leftrightarrow |-1\rangle$ as a two-level system with effective spin $s = 1/2$. In addition, the NV center interacts with nuclear spins, predominantly (abundance of 99.6\% \cite{felton2009hyperfine}) the intrinsic $^{14}$N with total spin $s = 1$, hyperfine coupling tensor $\bm{A_{\mathrm{N}}}$ and quadrupolar coupling $Q_{\mathrm{N}} \simeq -5.01$ MHz \cite{felton2009hyperfine}, causing further splitting of the $|\pm 1\rangle$ states. Due to the axial symmetry of the NV center, $\bm{A_{\mathrm{N}}}$ can be expressed in a diagonal form in the principal axis system of the NV axis with diagonal elements $\left[ A_{\mathrm{N}}^\bot \simeq -2.7\ \mathrm{MHz},\ A_{\mathrm{N}}^\bot,\ A_{\mathrm{N}}^\parallel \simeq -2.16 \mathrm{MHz} \right]$ \cite{felton2009hyperfine}.  Taking the above details into account, the relevant part of the NV center Hamiltonian can be written as 
\begin{equation}
		H_{\mathrm{NV}} = D_{\mathrm{ZFS}} J_z^2 + \gamma_e (\vec{B_0}\cdot\vec{J}) + \vec{J} \cdot \bm{A_{\mathrm{N}}} \cdot I_\mathrm{N}, 
\end{equation}
where $\gamma_e$ is the gyromagnetic ratio of an electron,  $\vec{J} = [J_x, J_y, J_z]$ and $\vec{I_{\mathrm{N}}} = [I_{\mathrm{N}x}, I_{\mathrm{N}y}, I_{\mathrm{N}z}]$ are, respectively, spin-operators of the NV center and $^{14}$N nuclear spin,  and $\vec{B_0} = [B_{0x}, B_{0y}, B_{0z}]$ is the vector of the externally applied magnetic field on the NV center.
The relatively large zero-field splitting allows to make the secular approximation $\left(|D_{\mathrm{ZFS}} + \gamma_e B_{0z}| >> A_{N}^\bot \right)$ where we can ignore all terms involving electron-nuclear spin flip-flops (containing $A_{\mathrm{N}}^\bot$). Further, we assume $B_{0z}>> B_{0x} \mathrm{\ and\ } B_{0y}$ and consequently choose $|0\rangle \leftrightarrow |+1\rangle$ (hereafter denoted $|1\rangle$) as two-level system. Now the NV center Hamiltonian simplifies to
\begin{equation}
		H_{\mathrm{NV}} =  (D_{\mathrm{ZFS}} + \gamma_e B_{0z}) J_z  + A_{N}^\parallel J_z I_{Nz}. 
\end{equation}
The RP consists of two radicals each containing an unpaired electron. A simplified reaction for the creation and recombination of radicals is shown in Fig.~\ref{setting}(b). The radicals can be generated either by electron transfer or chemical bond breaking from the oxidized state of the molecule. The reduction can be facilitated by various mechanisms, for example, by photoexcitation with light of appropriate wavelength (reduction of flavin based systems~\cite{schwinn2020uv}) or by chemical means (Haber-Weiss reaction \cite{kruszewski2003labile}). Depending on the internal molecular dynamics, the unpaired electrons in the radicals are born in either singlet ($s = 0$) $(|S_0\rangle  = [|\uparrow \downarrow\rangle - |\uparrow \downarrow\rangle]/\sqrt{2})$ or triplet ($s = 1$) $(|T_0\rangle  = [|\uparrow \downarrow\rangle + |\uparrow \downarrow\rangle]/\sqrt{2}, |T_+\rangle  = (|\uparrow \uparrow\rangle, |T_-\rangle  = (|\downarrow \downarrow\rangle)$ states, where $|\uparrow\rangle$ and $|\downarrow\rangle$ denote the eigenstate of the Pauli $z-$matrix. After creation, the singlet and triplet states inter-convert among each other under the RP Hamiltonian, which for the RP situated at location $(r, \zeta = \{\alpha, \beta\},$ see Fig.~\ref{setting}(a)) is given by~\cite{fay2020quantum}
\begin{align}
H_{\mathrm{RP}}(\zeta)  = &-\gamma_e [\vec{B}(\zeta).(\vec{S_1} + \vec{S_2})] -2J_{\mathrm{ex}} (\zeta) \vec{S_1}\cdot\vec{S_2} + \vec{S_1}.\bm{D}(\zeta).\vec{S_2} \nonumber\\
 &+ \sum_{i = 1}^{N_1} \vec{S_1}.\bm{A_{1i}}(\zeta).\vec{I_{1i}} +  \sum_{j = 1}^{N_2} \vec{S_2}.\bm{A_{2j}}(\zeta).\vec{I_{2j}},
 \label{RPHamilonian}
\end{align}
where $\vec{S_1}$ and $\vec{S_2}$ are the spin-operators of the unpaired electrons of the $1^{\mathrm{st}}$ and $2^{\mathrm{nd}}$ radical, respectively, and $\vec{B}(\zeta) $ denotes the external magnetic field applied on the RP. The radicals couple to each other via exchange interaction (coupling constant $J_{\mathrm{ex}} (\zeta)$) and dipolar interaction (coupling tensor $\bm{D}(\zeta)$). Each unpaired electron is further surrounded by a set of nuclei in the radical and interact with them via hyperfine coupling. The  $\vec{I_{1i}}(\bm{A_{1i}}(\hat{r}))$ and $\vec{I_{2j}}(\bm{A_{2j}}(\hat{r}))$ are spin-operators (hyperfine coupling tensors) of the $i^{\mathrm{th}}$ nuclei coupled to the unpaired electron in  the $1^{\mathrm{st}}$ and $j^{\mathrm{th}}$ nuclei coupled to the unpaired electron in the $2^{\mathrm{nd}}$ radical, respectively. The magnitude of hyperfine coupling is typically in the range of 2.8 - 28 MHz \cite{hiscock2016quantum} in organic radicals, whereas the strength of the dipolar and exchange interaction is dependent on the separation between the radicals \cite{efimova2008role}. Usually the dipolar and hyperfine coupling tensors are simulated or measured in a coordinate frame associated with the molecule that hosts the RP, hereafter referred to as the RP frame. The NV and RP frames are related by a rotation matrix $\mathcal{R}_\zeta$, i.e., $H_{\mathrm{RP}}(\zeta) = \mathcal{R}_\zeta^\dagger H_{\mathrm{RP}}'\mathcal{R}_\zeta$, where $H_{\mathrm{RP}}'$ is the RP Hamiltonian in the RP frame. 

To gain more insight into the inter-conversion dynamics of spin states of radicals, it is useful to work in the basis spanned by singlet and triplet states for the unpaired electrons. Using this choice of basis, we can divide $H_{\mathrm{RP}}(\zeta)$ into two parts $H_{\mathrm{RP}}(\zeta) = H_{\mathrm{RP}}^\mathrm{d} (\zeta)+ H_{\mathrm{RP}}^{\mathrm{nd}}(\zeta)$. The first part is diagonal in the singlet-triplet basis  for unpaired electrons irrespective of the basis chosen for the nuclei 
\begin{align}
 H_{\mathrm{RP}}^\mathrm{d} (\zeta) =& -\gamma_e [B_z(\zeta) (S_{1z} + S_{2z}) -2J_{\mathrm{ex}} (\zeta) \vec{S_1}\cdot\vec{S_2} \nonumber\\ &+ D_{s}(\zeta) (3 S_{1z} S_{2z} - \vec{S_1}.\vec{S_2}),
\end{align}  
where $D_s(\zeta) = -\frac{\mu_0 \gamma_e^2 \hbar}{4\pi |r_{RP}(\zeta)|^3}$ is the secular part of the dipolar coupling and $r_{RP}$ is the distance between the radicals. The second part $H_{\mathrm{RP}}^{\mathrm{nd}}(\zeta)$ contains terms corresponding to the transverse magnetic field, non-secular dipolar coupling contribution and hyperfine interactions with the surrounding nuclei. The singlet and triplet states are anti-symmetric and symmetric under spin exchange, respectively, and the evolution under $H_{\mathrm{RP}}^{\mathrm{nd}}(\zeta)$ breaks this  symmetry  which results in inter-conversion between these states. 

 Due to interaction with the surrounding environment, after the creation, along with the inter-conversion dynamics, the spin-correlated RP recombine back to the equilibrium state. The rates of recombination depend on whether the RP is in the singlet (rate constant $k_S$) or the triplet state (rate constant $k_T$). Products are thus formed with spin-state dependent yield and can be altered by the application of a suitable magnetic field, a process that is called the radical-pair mechanism (RPM)~\cite{hore2016radical}. The recombination dynamics can be modeled by treating the RP as an open quantum system with Lindbladian 
\begin{equation}
    \mathcal{L}_{rec}[{\cdot}]  = -\left\{\left(\frac{k_S}{2}|S_0\rangle\langle S_0| + \frac{k_T}{2} \sum_{i = +,0,-}|T_i\rangle\langle T_i|\right) \otimes I^{\otimes N}, \cdot \right\},
\end{equation}
Where $N = N_1 + N_2$ is total number of nuclei in the RP and $\left\{.\right\}$ denotes the anticommutator.

\begin{figure}[h]
	\begin{center}
		\includegraphics[trim=3.1cm 8cm 2cm 7.8cm,clip,width=8cm]{{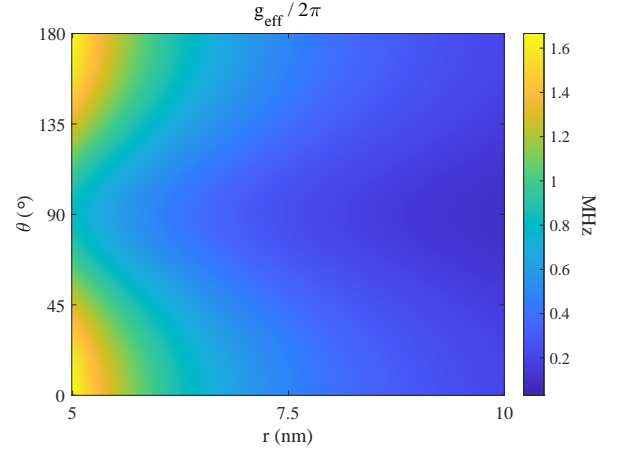}}
	\end{center}
	\caption{\small The effective coupling strength $g_{\mathrm{eff}}/ 2\pi =   D_r \sqrt{D_{cx}^2 + D_{cz}^2}/\pi$ for different relative distances $r$ between the RP and NV sensor and angles $\theta$ of the magnetic field  ($\phi = 0$ is assumed).}
	\label{effective coupling}
\end{figure}

The electron spin of the NV center and the RP is coupled via dipolar interaction and the corresponding Hamiltonian can be written as
\begin{equation}
		H_{\mathrm{C}}  = \vec{J}.\bm{D_c}.(\vec{S_1} + \vec{S_2}),
\end{equation}
where $\bm{D_c}$ is the dipolar coupling tensor in the NV frame which depends on the direction of the applied magnetic field $(\theta, \phi)$ and the NV centre distance to the RP. Under the secular approximation  $|D_{\mathrm{ZFS}} + \gamma_e B_z| >> |\bm{D_c}|$ and assumption $|B_z|> >\{|B_x|, |B_y|\}$, the coupling Hamiltonian simplifies to
\begin{equation}
		H_{\mathrm{C}} =  J_z D_{r} \sum_{i = x,y,z} D_{ci} (S_{1i} + S_{2i}) = J_z D_r \sum_{i = x,y,z} \Tilde{S_i} ,
\end{equation}

where $D_r = -\frac{\mu_0 \gamma_e^2 \hbar}{4\pi |r|^3}$, $D_{cx} = 3\sin2\theta\cos\phi/2$,  $D_{cy}  = 3\sin2\theta\sin\phi/2$, $D_{cz}  = 3\cos^2\theta - 1$ and $\Tilde{S_i} = D_{ci} (S_{1i} + S_{2i})$. The trace norm $g_{\mathrm{eff}} = ||D_r \sum_i \Tilde{S_i}||  = 2 D_r \sqrt{D_{cx}^2 + D_{cy}^2 + D_{cz}^2}$ represents the effective RP-NV coupling (Fig. \ref{effective coupling}) and can be determined using techniques developed in recent works for three-dimensional molecular localization using the NV center \cite{zhao2012sensing, zopes2018three, zopes2018three2,laraoui2015imaging}. Note that the coupling Hamiltonian can not be approximated to contain only the $J_z \tilde{S_z} $ term \cite{shi2015single, finkler2021quantum} as the  strength of cross terms ($J_z \tilde{S_x} $ and $J_z \tilde{S_y} $) may not be smaller compared to various other parameters in $H_{\mathrm{RP}}$ at low external magnetic fields.

Now, after establishing the general framework, the protocol to study RP dynamics using the NV center is as follows:
\begin{enumerate}
    \item Prepare the NV center and the RP in a known initial state $\rho(0) = \rho_{NV}(0)\otimes\rho_{RP}(0)$. The initial state of the unpaired electrons in the RP is either the $|S_0\rangle$ or $|T_0\rangle$ state as described before, while the nuclei start in the maximally mixed state $I^{\otimes N}/2^{N}$. The initial state of the NV center can be controlled and depends on the applied sensing scheme. 
     \item Solve the quantum dynamics in the NV frame using the master equation: $\frac{d\rho(t)}{dt} = -i[H, \rho(t)] - \mathcal{L}_{rec}[ \rho(t)]$. Since the NV center has the capability of detecting magnetic signal directly originating from inter-conversion dynamics of the RP instead of relying on the product yield based detection,  we can  simplify the Lindbladian by assuming $k_S = k_T = k$, thus giving $ \mathcal{L}_{rec}(\cdot) =  -k \left\{I^{\otimes (N+2)},\ \rho(t)\right\}$   \cite{fay2020quantum}. Further, the rate $k$ may be allowed to include decoherence of the electron spins of NV center as well because its typical time-scale is similar to recombination rates in organic RPs  which is in the order of tens of $\mu$s. 
    \item Trace out the RP to calculate the state of the NV center at any time $t$: $\rho_{NV}(t) = \mathrm{Tr}_{RP} [\rho(t)]$ and investigate signatures of the RP evolution on the states of the NV center.
\end{enumerate}
Depending on $r$, $\theta$ and $\phi$, the coupling between an NV center and an RP varies (Fig. \ref{effective coupling}). The resulting NV-RP dynamics can be divided in two dynamical regimes:
\begin{align}
    \mathrm{Weak \ coupling }: \ &g_{\mathrm{eff}} < \Gamma \\
        \mathrm{Strong \ coupling}: \ &g_{\mathrm{eff}}  > \Gamma,
\end{align}
    where $\Gamma = \frac{1}{\pi\ T_2}$ and $T_2$ is the dephasing time  of the NV center.  $T_2$ depends on factors including applied sensing sequence and magnetic field \cite{degen2017quantum}, proximity to surface impurities, presence of paramagnetic defects and nuclear spins around the NV center~\cite{doherty2013nitrogen}. Typical values of $T_2$ range from a few microseconds to a few milliseconds depending mainly on the applied sensing sequence, the isotopic purity of the diamond sample used \cite{mizuochi2009coherence} and the surface termination \cite{sangtawesin2019origins}, making both weak and strong coupling regimes achievable in practice.
\section{Simulations}
\label{sim}
In this section we present details and results of the simulations. We consider two RP systems, respectively with isotropic and anisotropic hyperfine coupling tensors. We investigate MFEs in RP systems in the weak as well as strong coupling regimes and discuss appropriate detection strategies. In the weak coupling regime, we study the MFE as a function of strength and direction of magnetic field due to a single as well as a small ensemble of RP. For the strong coupling regime, we point out potential challenges in the regime of single molecule detection that arise for larger bio-molecules.  
\subsection{RP systems}
The two RP systems we investigate are:
\begin{enumerate}
    \item Flavin adenine dinucleotide - Tryptophan (FAD$^{\bullet-}$ - TrPH$^{\bullet+}$) with anisotropic hyperfine couplings, which has been subjected to various MFE studies using absorption spectroscopy in recent years~\cite{evans2013magnetic}. This system is considered to be the RP behind avian magnetoreception~\cite{hore2016radical}. 
    \item Pyrene-N,N-dimethylaniline (PY$^{\bullet-}$ - DMA$^{\bullet+}$) RP, widely studied \cite{werner1977theory} and with isotropic hyperfine couplings \cite{rodgers2007determination}.   
\end{enumerate}

\subsection{Weak coupling regime}
In the weak coupling regime, no observable splitting of the NV $|1\rangle$ level occurs upon interaction with the RP and the effect of spin-dynamics of the RP on the NV center dynamics then appears only as a time-dependent classical magnetic field.   The coupling Hamiltonian simplifies to
	\begin{align}
		H_{\mathrm{C}} (\zeta, t) & = J_z D_r \ \sum_{i = x,y,z} \langle \Tilde{S_i}( \zeta, t) \rangle,
		\label{classical}
	\end{align}   
	where $\langle O (t) \rangle = \mathrm{Tr} [O\  e^{-i H_{\mathrm{RP}} (\zeta)t}\ \rho_{RP} (0)\ e^{i H_{\mathrm{RP}}(\zeta)t}]$ with $O \in \{\Tilde{S_i}\}, i = x, y, z$. The detectable magnetic signal generated from the RP  situated at a location $(r, \zeta)$ can now be calculated as (in units of Tesla):
	\begin{align}
	X_i(t, \zeta) &= \xi \int_{0}^{2\pi}\int_{0}^{\pi/2}\int_{r_1}^{r_2} \frac{D_r}{\gamma_e} \langle \Tilde{S_i}(\zeta, t) \rangle \ |r|^2\ \sin\alpha\ dr\ d\alpha\ d\beta \nonumber\\
		&= \frac{\xi\ \mu_0 \gamma_e\ \hbar}{2} \log \left(\frac{r_2}{r_1} \right)\int_{0}^{\pi/2}  \langle \Tilde{S_i}(\zeta, t) \rangle\ \sin\alpha\ d\alpha\ ,
	\end{align}
	where $i = x, y, z$, $\xi$ is the number density of RPs, and $(r_2-r_1)$ is the sensing radius of the NV center (see Fig.~\ref{setting}(a)). The maximum signal is generated when $\mathcal{R}_\zeta = \mathcal{I}$, where $\mathcal{I}$ is the identity matrix, i.e. when the RP frame is aligned with the NV frame:
	\begin{equation}
	   X_i^{max}(t) = \frac{\xi\ \mu_0 \gamma_e\ \hbar}{2} \log \left(\frac{r_2}{r_1}\right) \langle \Tilde{S_i}(t) \rangle. 
	\end{equation}
	The possibility of achieving such a condition is discussed in Appendix \ref{appendix A}. 
	
	\begin{figure}[b]
	\begin{center}
		\includegraphics[trim=11.1cm 9cm 8.4cm 3cm,clip,width=8.5cm]{{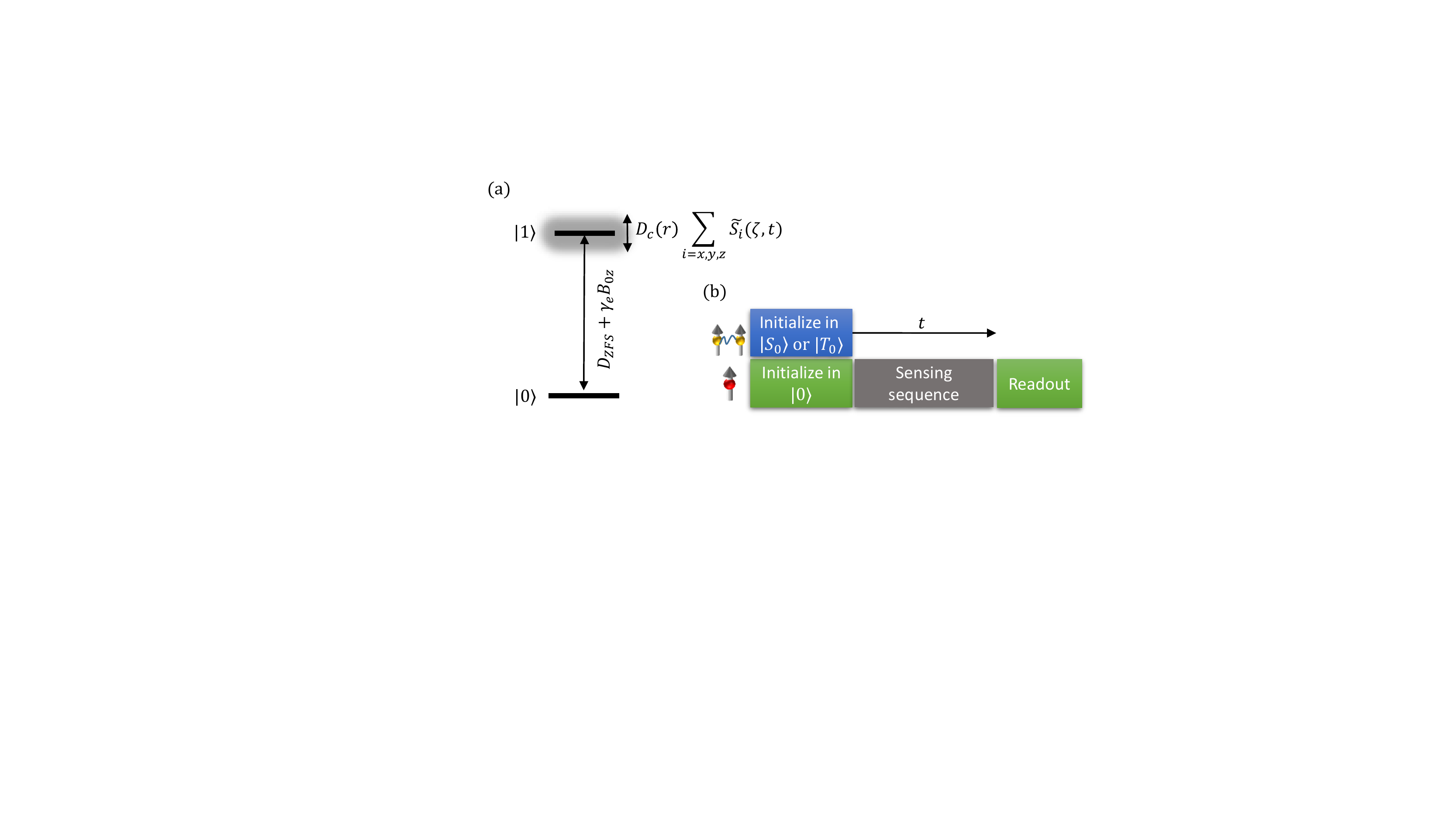}}
	\end{center}
	\caption{\small  Energy level structure of an NV center in  the weak coupling regime. The RP causes a relative phases between $|0\rangle$ and $|1\rangle$ states of NV center that is proportional to $H_C(\zeta, t)$.  The sequences to detect the RP dynamics in each regime are shown in (b). The RP is born in $|S_0\rangle$ or $|T_0\rangle$ state and the NV is initialized in $|0\rangle$ state. Then an appropriate sensing sequence is applied which is chosen based on the frequency profile of $X_i(\omega, \zeta)$. During sensing, the NV center acquires a phase which contains the information about the RP dynamics which is then read out by a projective measurement. }
	\label{weak_coupling_level}
    \end{figure}
	
	A simple protocol to detect the above signal is shown in Fig.~\ref{setting}(d). First, we initialize the NV center in $|0\rangle$ and the RP in either the $|S_0\rangle$ or $|T_0\rangle$ state. An appropriate sensing sequence is then applied on the NV. The choice depends on the Fourier spectrum of the detectable signal, denoted by  $X_i(\omega, \zeta)$ here. Finally, a projective measurement is made on the NV to yield information about the accumulated phase, which is proportional to $X_i(t, \zeta)$, acquired during the sensing sequence. 
	
	As evident from Eq.~(\ref{classical}), only specific combinations of RP density matrix elements corresponding to  $\{S_{ji}, j = 1,2\ \mathrm{and}\ i = x, y, z\}$ contribute to $X(t)$ and are directly measurable by the NV center.  Explicitly, in the singlet-triplet basis, the directly measurable elements are: 
\begin{align}
 S_{1z} + S_{2z} & = |T_+\rangle\langle T_+| - |T_-\rangle\langle T_-|, 
 \label{Sz}
\end{align}
which measures the magnetization corresponding to the population difference of outer triplet states and
\begin{align}
S_{1x} + S_{2x} &= \sqrt{2}\ \mathrm{Re} (|T_+\rangle\langle T_0| + |T_-\rangle\langle T_0|) \nonumber\\
S_{1y} + S_{2y} &= \sqrt{2}\ \mathrm{Im} (|T_+\rangle\langle T_0| + |T_-\rangle\langle T_0|),
\label{Sxy}
\end{align}
which measure the magnetization corresponding coherences between the outer and central triplet states. Thus the detection of magnetization corresponding to elements $S_{1x} + S_{2x}$ and $S_{1y} + S_{2y}$ can reveal involvement of quantum coherent phenomenon in RP dynamics in singlet-triplet basis. The full state tomography of  density matrix of RP, however, requires conversion of various elements onto directly detectable ones (Eq. \ref{Sz} and \ref{Sxy}) and is beyond the scope of this work. 
	

We consider RPs to be statistically distributed above the surface of the diamond sample. The coupling strength of an RP outside the sensing radius of three-four times the depth $d$ of the NV centre (cf. Fig.\ref{setting}(a)) decreases by an order of magnitude compared to RPs on the surface. The number of RPs falling in the sensing volume depends on the size of the host protein. For the case of proteins with many amino acids and large molecular weights (>$500k$Da, the average radius $>5$nm \cite{erickson2009size}), only one of them might fall in the sensing volume with high probability. On the other hand, for the case of smaller proteins (<50kDa, the average radius being 2-3 nm), tens of RPs can be accommodated in the sensing volume. The following two subsections discuss how the generated signal behaves in these two cases. 

\subsubsection{RP on single molecule}
To study RP dynamics at the single-molecule level, we assume $\mathcal{R}_\zeta = I$ (therefore we  drop $\zeta$ notation for discussion in this subsection) and $\alpha = \beta = 0$ without loss of generality. We use three nuclei with maximum anisotropic hyperfine couplings for each of the radicals in (FAD$^{\bullet-}$ - TrPH$^{\bullet+}$) : $\mathrm{N_5, N_{10}}$, and $\mathrm{H_6}$ for FAD$^{\bullet-}$, and $\mathrm{N_1, H_1}$, and $\mathrm{H_{\beta 1}}$ for TrPH$^{\bullet+}$)~\cite{hiscock2016quantum}. The used $J_{\mathrm{ex}}$ and $\bm{D}$ couplings correspond to FAD$^{\bullet-}$ - TrPH$^{\bullet+}$ RP in \textit{Drosophila melanogaster} cryptochrome protein~\cite{fay2020quantum}. The recombination rate is assumed to be $k = 2 \times 10^5$ Hz which is of the order of the observed spin relaxation time in behavioural studies in migratory birds \cite{kobylkov2019electromagnetic}. To calculate the total signal received (or phase accumulated by the NV center) at a given strength and direction of applied magnetic field, we define the time integrated signal per second as
\begin{equation}
    X_i^I  = \frac{1}{T}\sum_t X_i (t).
\end{equation}

\begin{figure}[t]
	\begin{center}
		\includegraphics[trim=8.2cm 0.6cm 7cm 0cm,clip,width=9.5cm]{{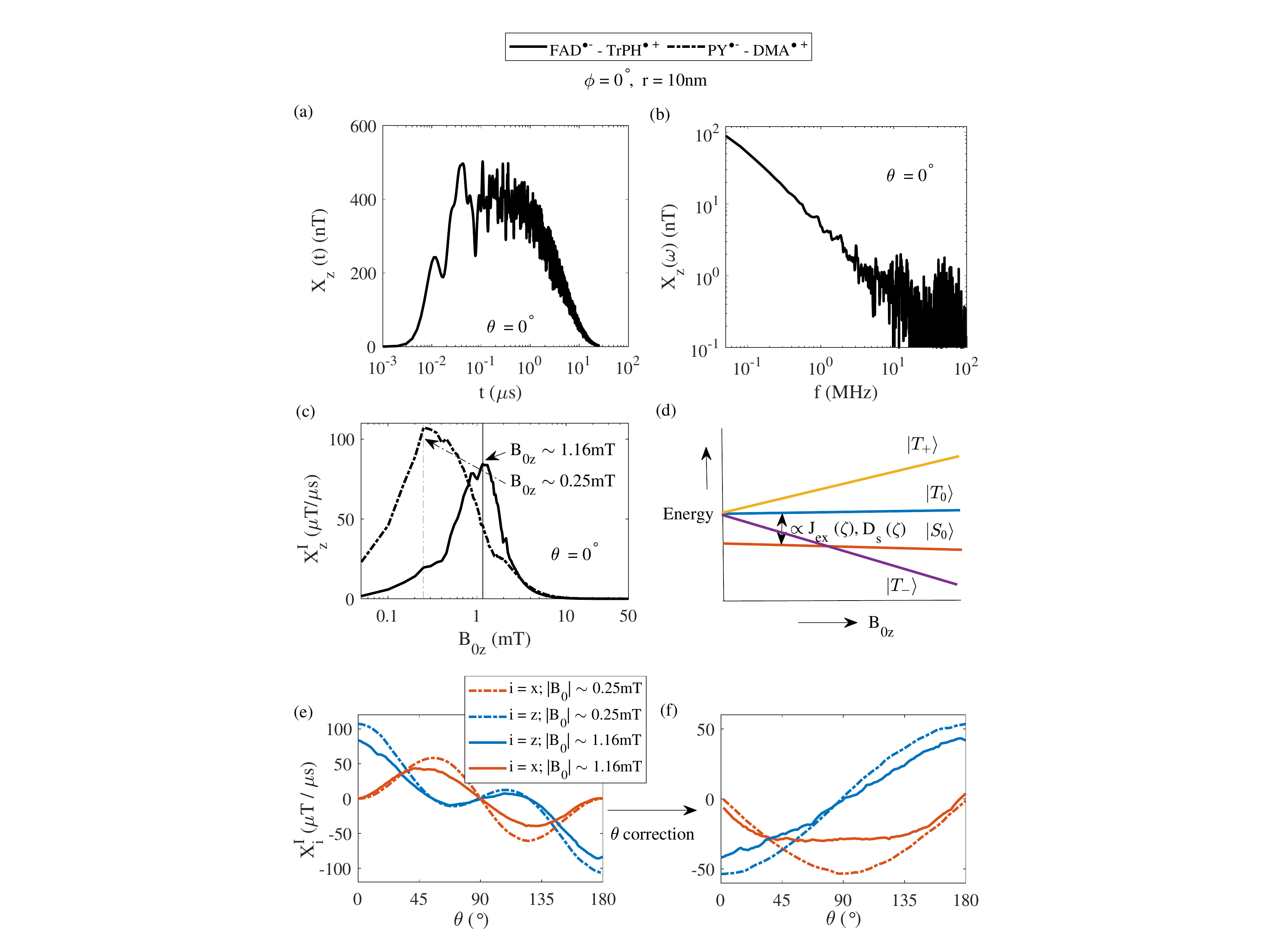}}
	\end{center}
	\caption{\small Signal received by the NV center from a single RP molecule of model systems FAD$^{\bullet-}$ - TrPH$^{\bullet+}$ (solid lines) and PY$^{\bullet-}$ - DMA$^{\bullet+}$ RP) (dashed lines).  Time (a) and frequency (b) profile of the signal at magnetic field $B_{0z} = 1.16$mT for FAD$^{\bullet-}$ - TrPH$^{\bullet+}$ RP, where maximum of time integrated signal $X_z^I$ and $X_x^I$ is obtained as plotted in (c). $\theta = 0^\circ$ is assumed in (a), (b), and (c). (d) Level dynamics of the RP in singlet-triplet basis as function of the strength  of the applied magnetic field. Dependence of the integrated signal on the angle of the applied magnetic field with (e) and without (f) $\theta$ correction at magnetic field $|B_{0}| = 1.16$mT and |$B_{0}| = 0.25$mT, respectively, for FAD$^{\bullet-}$ - TrPH$^{\bullet+}$ and PY$^{\bullet-}$ - DMA$^{\bullet+}$ RP. }      
	\label{weak}
\end{figure}

In Fig.~\ref{weak}, we show the strength of the magnetic signal received by the NV from an RP on a single molecule at a distance $r = 10$nm. For such shallow NV centres, the achievable sensitivity is of the order of a few $\mathrm{nT/\sqrt{Hz}}$ to a slowly alternating signal \cite{shi2015single, grinolds2013nanoscale}. For the case of $\phi = 0$, as we assume here, the detectable signal is only generated by $X_x^I $ and $X_z^I$, which, respectively, corresponds to the dynamics of RP density matrix elements $S_{1x} + S_{2x}$ and $S_{1z} + S_{2z}$. 

 To determine the suitable NV sensing sequence to probe the generated signal, Fig.~\ref{weak}(a) plots the temporal evolution of $X_z(t)$ for a bias magnetic field that maximizes the time integrated signal $X_z^I $ (see Fig.~\ref{weak}(c)). The unpaired electrons in the RP are initialized in the singlet-state whose population decays with time as it is converted into other density matrix elements under the evolution of $H_{\mathrm{RP}}$, thus resulting in the build-up of $X_z^I$. The generated signal (hundreds of nT) is within the sensitivity achieved for $<10$nm deep NV centres. The corresponding spectrum is shown in Fig.~\ref{weak}(b). Because we are interested in low frequency dynamics (kHz to tens of MHz) resulting from the dipolar and hyperfine coupling terms (typically 1-10mT) in $H_{\mathrm{RP}}$, the maximum component of the signal in the frequency domain is limited to 500 MHz. The cut-off frequency of these oscillations observed is in the order of 10MHz. Sequences including Ramsey, spin-echo, and dynamical decoupling are thus suitable for probing such RP dynamics~\cite{degen2017quantum}.

\textit{Variation with strength of the magnetic field} $-$ To study the dependence of the signal on the strength of the applied magnetic field, we set $\theta = 0$, i.e., $B_{0x}=0$ and $B_{0y}=0$ so that there is no spin mixing of the NV center states keeping the two-level approximation valid  and $B_z$ is varied from 0 to 50mT. The dependence of the time-integrated signal $X_z^I$ $\left(X_x^I = X_y^I = 0\ \mathrm{in\ this\ case\ as}\ \theta = 0\right)$ is plotted as a function of $B_z$ in Fig.~\ref{weak}(c). 

The features of the time-integrated signal $X_z^I$ can be qualitatively explained by the dynamics of the RP under $H_{\mathrm{RP}}^{\mathrm{nd}}$ in singlet-triplet basis as shown in Fig.~\ref{weak}(d).  At $B_{0z} \approx 0$, all three triplet states are almost degenerate and the energy gap between $|S_0\rangle$ and $|T_0\rangle$ states is proportional to $J_{\mathrm{ex}}$ and $D_s$. With the unpaired electrons in the RP initialized in the singlet state, singlet-triplet oscillations occur due to evolution of the RP under $H_{\mathrm{RP}}^{\mathrm{nd}}$ containing both a non-secular dipolar coupling contribution and hyperfine interactions with the surrounding nuclei (Eq. \ref{RPHamilonian}). However, for small $B_{0z}$, the RP only generates a very low signal $X_z^I$ as outer triplet states ($|T_\pm\rangle$) remain close to degeneracy. As $B_{0z}$ increases, the outer triplet states  move apart and give rise to additional manifolds for singlet-triplet mixing \cite{brocklehurst1976spin, brocklehurst1996free}.  This opens the possibility of information (population and coherence) transfer among $|S_0\rangle$ and $|T_0\rangle$ states to only one of the outer triplet states, creating a population imbalance between outer triplet  states and giving rise to an increase in $X_z^I $. At a certain $B_{0z}$ value, there is a maximum transfer  between singlet-triplet oscillations to population difference of outer triplet states, giving rise to the peak-like features in $X_z^I$ (for  example the peak at $\sim  1.16$ mT and $\sim 0.25$ mT, respectively, for FAD$^{\bullet-}$ - TrPH$^{\bullet+}$ RP and PY$^{\bullet-}$ - DMA$^{\bullet+}$ RP). This phenomenon, which is closely related to low field effects (LFE) in the context of singlet yield \cite{lewis2018low}, occurs due to  level crossing between $|S_0\rangle$ and either of the outer triplet \cite{till1997radical} states. A further increase in $B_{0z}$ results in energetic isolation of the  outer triplet states which causes steady decrease in the probability of information transfer between them, and $|S_0\rangle$ and $|T_0\rangle$ states, eventually vanishing at very high magnetic fields $|B_{0z}|>> |H_{\mathrm{RP}}^{\mathrm{nd}}|/\gamma_e$.

\textit{Variation with direction of the magnetic field} $-$ In Fig. \ref{weak}(e), we investigate the signal dependence on the direction of the applied magnetic field. Here we restrict the magnetic field to the $XZ$ plane by assuming $\phi = 0$ as the following discussion holds for the general case. Although $B_{0x}$ is nonzero for certain values of $\theta$, its magnitude should be less than that required for mixing of the $|1\rangle$ and $|-1\rangle$ levels of the NV center. We fix the magnetic field strength to $\sim  1.16$ mT and $\sim 0.25$ mT, respectively, for FAD$^{\bullet-}$ - TrPH$^{\bullet+}$ RP and PY$^{\bullet-}$ - DMA$^{\bullet+}$ RP where the maximum MFE is expected based on the results shown in Fig.~\ref{weak}(c). The dependence of the time-integrated signal on $\theta$ is plotted in Fig.~\ref{weak}(e). The shape of the signal received $X_x^I$ and $X_z^I $ ($X_y^I = 0$ since $\phi = 0$)  is dominated by the shape of $D_{cx}$ and $D_{cz}$, respectively. To reveal the true dynamics of the RP with $\theta$, we calculate the normalized signals instead in Fig. \ref{weak}(f). 

Once again, using the level dynamics under $H_{\mathrm{RP}}^{\mathrm{nd}}$ in the singlet-triplet basis, a qualitative explanation of various features of the generated signals is possible using the following arguments: (i) for a given magnetic field, $\theta$ close to $90^\circ$ essentially means presence of an extra channel for singlet-triplet mixing due to nonzero $B_{0x}$ and $B_{0y}$ and (ii) at $\theta$ close to $90^\circ$, the outer triplet states are almost degenerate with maximum splitting occurring at $\theta$ close to $0^\circ$ or $180^\circ$. The shape of $X_x^I$ is a direct consequence of argument (i) as nonzero transverse fields close to $90^\circ$ increase the probability of singlet-triplet as well as triplet-triplet mixing. The signal $X_z^I$ is a consequence of argument (ii) as higher splitting of outer triplet states opens new manifolds for information transfer between them and $|S_0\rangle$ and $|T_0\rangle$ states as described earlier.

For parameter values chosen here and described above, we do not observe the recently discovered spike feature~\cite{hiscock2016quantum} in the generated signals $X_z^I$ and $X_x^I$ presumably due to the high value of  magnetic field used, and the non-zero exchange and dipolar coupling in the $H_{\mathrm{RP}}$ \cite{hiscock2017disruption}. However, the spike  feature  can be observed when we consider a simpler RP system with only one nuclear spin in each RP in earth's magnetic field  as summarized in Appendix~\ref{appendix B}. 

\subsubsection{Ensemble of RPs}
\begin{figure}[h]
	\begin{center}
		\includegraphics[trim=2.8cm 4.3cm 3.1cm 3cm,clip,width=8cm]{{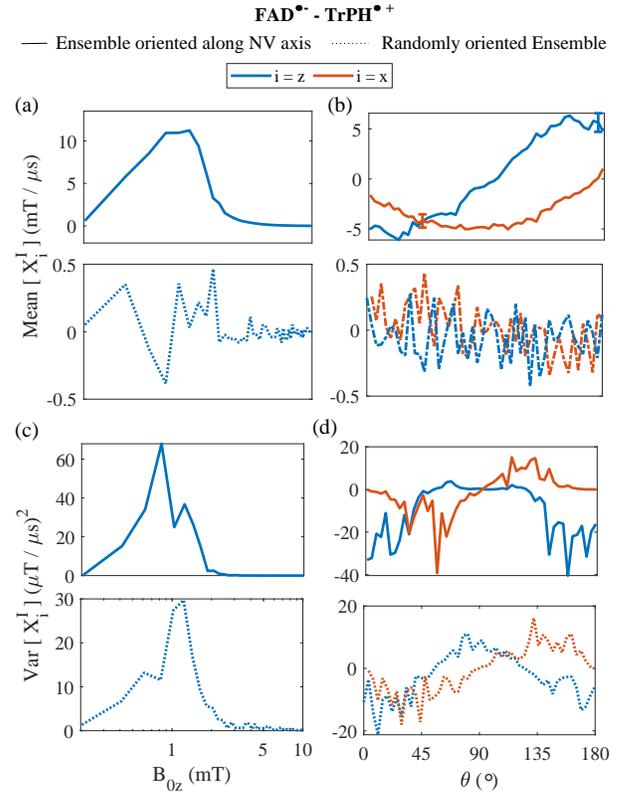}}
	\end{center}
	\caption{\small Dynamics of the statistical signal generated by an ensemble of RPs in the sensing volume of the NV center. To reduce computation time, two nuclear spins were included for each radical: $\mathrm{N_5, N_{10}}$ for FAD$^{\bullet-}$, and $\mathrm{N_1, H_1}$ for TrPH$^{\bullet+}$). MFE against variation in (a and c) strength of magnetic field (b and d) direction of the magnetic field ($\theta$ corrected).  The solid line indicate  the average MFE when the coordinate frames of all the RPs are aligned ( $\mathcal{R}_\zeta = I$) with the NV frame, while the dashed lines show the average MFE from a  randomly oriented ensemble (mean and variance of 50 random realizations of $\mathcal{R}_\zeta$).}
	\label{weak ensemble}
\end{figure}

Assuming the minimum size of the protein to be 3-4 nm (small protein with molecular weight $<50$kDa~\cite{erickson2009size}), $\approx 100$ RPs ($\xi = 5 \times 10^{-2}/\mathrm{nm}^3$) might fall in the NV sensing volume. In Fig.~\ref{weak ensemble} we plot the mean  and variance of the generated signal as a function of strength and direction of applied magnetic field for an ensemble of RPs. We consider two types of  ensembles: (i) All the RPs are oriented along the direction of the NV axis, i.e., $\mathcal{R}_\zeta = I$ (solid lines in Fig .\ref{weak ensemble}), and (ii) all RPs are randomly oriented with respect to the NV axis (dashed lines in Fig.~\ref{weak ensemble}), i.e. $\mathcal{R}_\zeta = R_\zeta^x(\alpha)R_\zeta^y(\beta)R_\zeta^z(\gamma)$, where $R_\zeta^x(\alpha)$, $R_\zeta^y(\beta)$, and $R_\zeta^z(\gamma)$, respectively, are rotation matrices about the $x$, $y$ and $z$ axes with randomly chosen Euler angles $\alpha$, $\beta$, and $\gamma$. For both cases, the distance between RPs and the NV were sampled from a uniform distribution in the range $5 - 20 \mathrm{nm}$. 

For the uniformly oriented ensemble (i), the mean of the generated signal is obviously amplified in contrast to the randomly oriented ensemble in (ii) where the features due to the RP spin dynamics are almost entirely removed because of random averaging of the signal over various realizations of $\mathcal{R}_{\zeta}$. The variance of the generated signal as function of strength of the magnetic field is within the sensitivity range of a shallow NV center for both types of ensembles and the LFE behaviour akin to the single-molecule case (Fig. \ref{weak} (c)) is still observable after averaging. On the contrary, although there is an observable signal when $\theta$ is varied, the shape of the variation is erratic with no similarity to the single molecule case (Fig. \ref{weak} (f)). Therefore, although the ensemble (ii) is simply realized by drop casting a molecule sample on the surface of the diamond, detecting a MFE in this arrangement is more challenging. The situation might be compared to absorption or EPR studies where the size of the probed ensemble is very large (micro litres of sample) and hence MFE detection relies on statistical signals generated from a very small number of molecules. This limits the detection of MFE to a few percent~\cite{maeda2012magnetically} along with causing wastage of precious biological samples. Hence, it becomes important to appropriately position the proteins in the desired orientation with respect to the NV center to maximize the probability of successful detection of an MFE (Appendix~\ref{appendix A}).

\subsection{Strong coupling regime}
\begin{figure}[h]
	\begin{center}
		\includegraphics[trim=10.6cm 1.4cm 10.5cm 2cm,clip,width=8.2cm]{{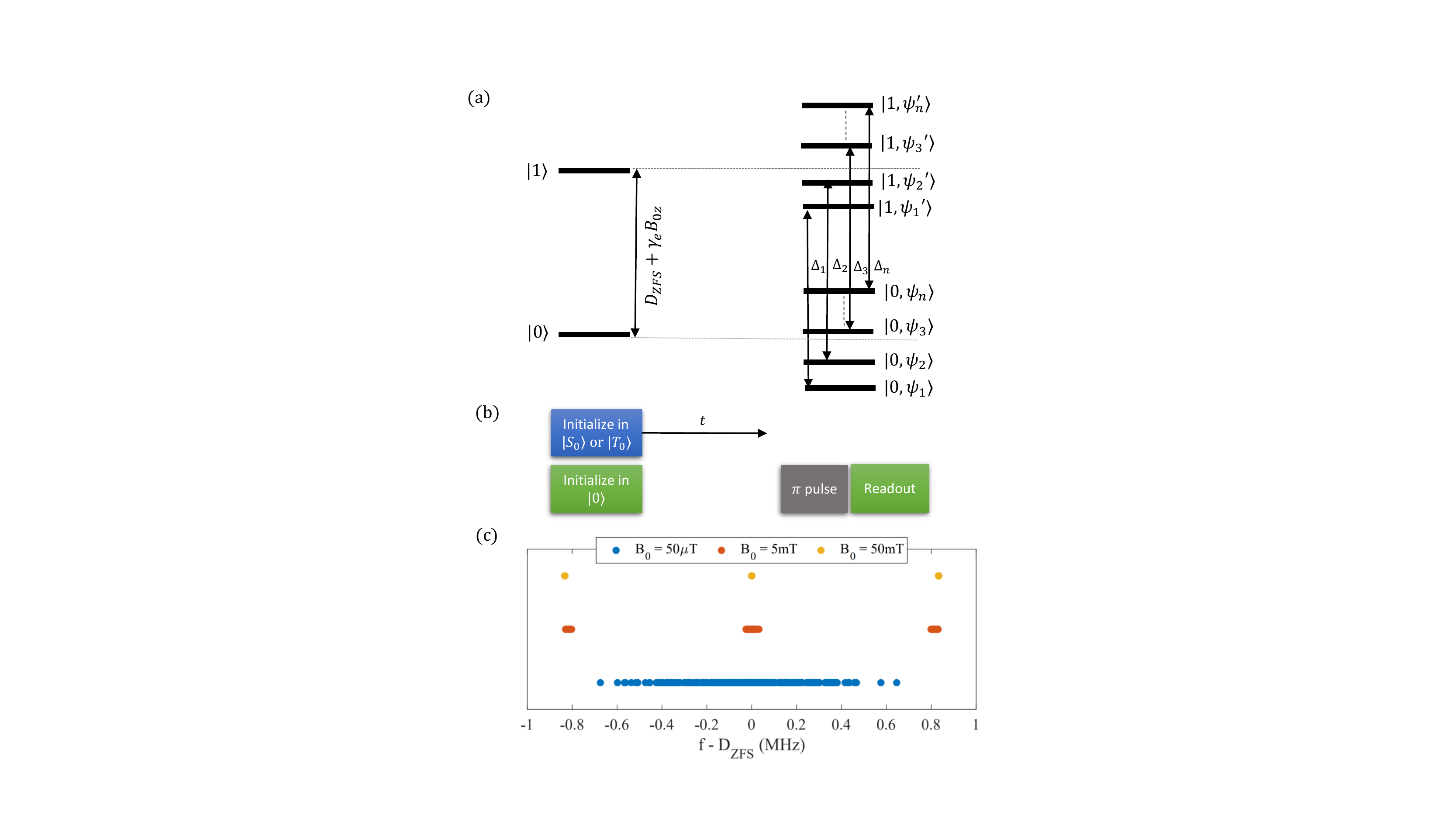}}
	\end{center}
	\caption{\small  (a) Energy level structure in the strong coupling regime. The RP components of an eigenstate in $|0\rangle$ manifold ($|\psi_i\rangle$) are determined by $H_{\mathrm{RP}}$ alone whereas in the eigenstate $|1\rangle$ manifold, ($|\psi_i'\rangle$) are determined by both $H_{\mathrm{RP}}$ and $H_c$ together. Here $\{(\Delta_i'-\Delta_i)\}$ depend on the  strength of the coupling tensor $\bm{D}$. The sequence to detect RP dynamics in this regime is shown in (b). (c) The number (each circle represent one peak) of peaks in spin resonance spectrum of NV center resulting from interaction of the NV center with an RP as a function of applied magnetic field for a single RP. }
	\label{strong_coupling_level}
\end{figure}

 In the strong coupling regime, the energy levels of the NV and RP mix, giving rise to a level structure as illustrated in Fig.~\ref{strong_coupling_level}. Due to the $s = 1$ structure of the NV center, the RP components of eigenstates in $|0\rangle$ and $|1\rangle $ are quantized along different axes determined by the strength of the coupling tensor. The total Hamiltonian in this regime can be reduced to 
	\begin{align}
	   	H &= H_{\mathrm{NV}} + H_{\mathrm{RP}} (\zeta) |0\rangle\langle0| +  \left[H_{\mathrm{RP}}(\zeta) + D_r \sum_{i = x,y,z} \Tilde{S_i}\right]|1\rangle\langle 1|.
	\end{align}

Depending on the magnitude of various couplings in $H_{\mathrm{RP}} (\zeta)$, the $|0\rangle$ and $|1\rangle$ states of the NV center are further split into a maximum of $2^{N+2}$ levels, as shown in Fig.~\ref{strong_coupling_level} (a). The level structure amounts to at maximum $2^{N + 2}$ resonances in the  magnetic resonance spectrum of the NV center, corresponding to transitions $|+1, \psi'_n\rangle\rightarrow |0, \psi_n\rangle$, where $|\psi'_n\rangle$ and $|\psi_n\rangle$ are respectively eigenstates of $ H_{\mathrm{RP}} (\zeta)$ and $ H_{\mathrm{RP}} (\zeta) + D_r \sum_i \Tilde{S_i}$, and  $n = 1$ to  $N + 2$. Monitoring the magnitude of these resonance peaks may facilitate a probing of the dynamics of the populations of $ H_{\mathrm{RP}} (\zeta)$ using a simple sequence as shown in Fig. \ref{strong_coupling_level} (b).
	
	The magnitude of the spin resonance peaks is proportional to the population difference of $|\psi'_n\rangle$ and $|\psi_n\rangle$ states. Taking the contribution from all RPs into account, it can be calculated as $C_n(\zeta, t) = $
	\begin{align}
	    \xi\int_{0}^{2\pi}\int_{0}^{\pi/2}\int_{r_1}^{r_2} \sum_{i = 1}^n  \left[\langle P_{\psi_n'} \rangle (\zeta, t) -  \langle P_{\psi_n} \rangle (\zeta, t)\right] |r|^2\ \sin\alpha\ dr\ d\alpha\ d\beta,
	\end{align}
	where $P_{\psi_i} = |\psi_i\rangle\langle\psi_i|$ is the projector corresponding to the state $|\psi_i\rangle$. There are a couple of important details about the detection of RP dynamics in this regime which should be pointed out: 
	\begin{enumerate}[label=(\roman*)]
	
	\item  Since we wish to study RP dynamics exclusively under an applied magnetic field, a pulsed sensing sequence (Fig \ref{strong_coupling_level} (b)) is preferred as it ensures that the NV center stays in the $|0\rangle$ state during the time of evolution, implying no backaction of the NV on the RP. In a Ramsay type of sequence where the NV center is prepared in $(|0\rangle + |1\rangle) /\sqrt{2}$, the RP experiences an additional induced field from the $|1\rangle$ state component \cite{finkler2021quantum} which is different for each RP, complicating the probing of MFEs. 
	
	    \item As the size of the bio-molecule increases, the number of unavoidable hyperfine interactions also grows. As a consequence, the spectral features will start to overlap, especially when the applied magnetic fields are comparable to the hyperfine couplings in $H_{\mathrm{RP}} (\zeta)$ (see Fig. \ref{strong_coupling_level} (c) for FAD$^{\bullet-}$ - TrPH$^{\bullet+}$). In this case it becomes challenging to access the various electronic transitions of the RP separately. One could consider decoupling the nuclear spin bath \cite{de2012controlling, bauch2018ultralong} during the readout time by driving the RP at a frequency larger than the strongest hyperfine coupling, however the short readout duration (typically $\approx 300$ns for NV ceneter \cite{doherty2013nitrogen, steiner2010universal}) will render the averaging  inefficient. 
	\end{enumerate}
In the light  of the second argument, although the detection of RP dynamics in the strong coupling regime may be challenging, it offers an opportunity to track various populations of the RP density matrix individually and the possibility of quantum control of the RP using an NV center \cite{finkler2021quantum}. 

\section{Conclusion and Outlook}
 \label{conclude}
 In summary, we analysed the realistic prospects of MFE detection at the single molecule scale in RP reactions in bio-molecules using a single NV center in diamond. To realistically describe the bio-molecular spin dynamics, our RP model includes three nuclear spins per radical that have the largest hyperfine couplings. Including dipolar and exchange coupling between the two electrons forming the radical pair, the RP spin dynamics becomes correlated and can not be simulated by treating each of the radicals independently ~\cite{liu2017scheme}. Depending on distance and orientation between NV and RP, two coupling regimes of dynamics can be considered requiring accordingly tailored NV sensing approaches. We find that in the weak coupling regime, RP dynamics can be seen as a classical magnetic field by the NV. This regime is most suitable for large bio-molecules with unavoidable and spectrally indistinguishable spin
interactions. The signal generated even from a single RP is well within the sensitivity achievable by state of the art NV centers and distinct features of RP spin dynamics are thus observable at the single molecule and a small ensemble ($\leq$ 100 molecules) scale.  

The simulations performed in our work can be further improved by incorporating more nuclear spins and using the recently developed coherent state sampling method \cite{lewis2016efficient, fay2017spin} for efficient computation of spin dynamics. Adding to the recent theoretical works \cite{ikeya2019single, liu2017scheme, finkler2021quantum}, we expect our analysis to pave the way towards experimental detection of MFE in bio-molecules at the single molecule level and unravel the importance of quantum coherent effects in biochemical processes~\cite{kim2021quantum, marais2018future}. 

\section{Acknowledgments}
We acknowledge financial support from the Novo Nordisk Foundation through the projects bioQ, bio-mag, and QuantBioEng, the Danish National Research Foundation (DNRF) through the center for Macroscopic Quantum States (bigQ, Grant No. DNRF0142), the Villum Foundation through the project bioCompass, and the EMPIR program co-financed by the Participating States and from the European Union’s Horizon 2020 research and innovation programme via the QADeT project (Grant No.
20IND05).

\section{Appendix}
\subsection{Orientation of RP with respect to NV}
As described in the main text, the orientation of the molecule hosting the RP is crucial for obtaining maximum signal.  Recent studies involving investigation of structure and motion of single proteins using NV centers have employed either statistical placement of proteins via diamond surface functionalization \cite{lovchinsky2016nuclear, shi2015single, sushkov2014all} or controlled alignment via attaching the protein to a solid host and then using a nanopositioning system, for example a tuning fork scanning probe of an atomic force microscope \cite{wang2019nanoscale}. The latter method is preferred because, in the former case, there are two main disadvantages: (i) The RP might end up in an orientation where one or more coupling parameters ($D_{ci}, i = x, y, z$) are zero, thereby reducing the coupling to the NV center (ii) The functionalization of the diamond surface may not be possible for all types of proteins.    
\label{appendix A}

\subsection{Spike  features in the generated signal}
\label{appendix B}
Here we consider a simple RP model system with only one nuclear spin in each radical, with hyperfine tensors $\bm{A_1}$ and $\bm{A_2}$, to analyze which parameter ranges give rise to 
spike-like features of the generated signal as a function of the direction of the applied magnetic field. This feature has been suggested to be behind the precision of an avian magnetic compass \cite{hiscock2016quantum}.   

\begin{figure}
	\begin{center}
		\includegraphics[trim=3.1cm 5.3cm 2.7cm 5.8cm,clip,width=8.5cm]{{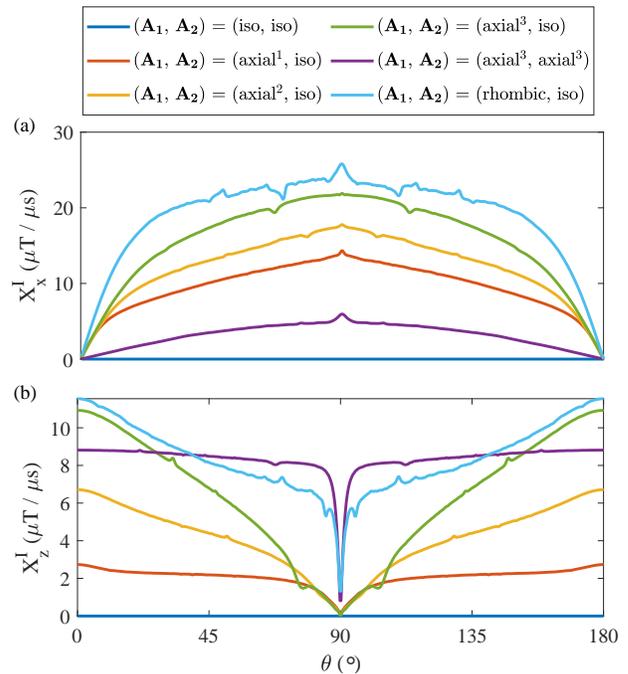}}
	\end{center}
	\caption{\small The generated signal $X_x^I$ and $X_z^I$ for various types of hyperfine coupling tensors for the model system. Here case `iso', `axial$^1$, `axial$^2$', `axial$^3$', and `rhombic', stands for hyperfine principle components $(A_{xx}, A_{yy}, A_{zz})$, respectively,  $ (0.5, 0.5, 0.5)$mT, $(-0.09,-0.09 , 1.76)$mT, $(-0.2, -0.2 , 1.76)$mT, $(-0.39, -0.39 , 1.76)$mT, and $(-0.39, 0 , 1.76)$mT.  The exchange coupling is chosen to be $J_{\mathrm{ex}} = 0.25$mT and diplolar coupling tensor ($\bm{D}$) is assumed to be zero as the effect is similar to exchange coupling. The lifetime of the RP is $5\mu$s.}
	\label{hyp_vary}
\end{figure}

In Fig  \ref{hyp_vary}, we plot the generated signal  $X_x^I$ and $X_z^I$, respectively, in (a) and (b) for various types of hyperfine coupling tensors chosen by varying principal axis components set $(A_{xx}, A_{yy}, A_{zz})$. For the anisotropic hyperfine coupling case, the principal axis system is that of the $\mathrm{N_5}$ in FAD$^{\bullet-}$ \cite{hiscock2016quantum}. A qualitative description of the seen behaviour can be provided by looking at the symmetry of the RP Hamiltonian. In the absence of dipolar coupling, there are three interactions in the RP Hamiltonian, namely, Zeeman, exchange and hyperfine. Zeeman (same magnetic field on all the spins) and exchange interactions are symmetrical with respect to electronic spin exchange while hyperfine interaction is symmetry breaking in general. When the hyperfine interaction is also assumed to be isotropic, the full Hamiltonian is symmetric, and singlet and triplet are eigenstates. As a result, there is no singlet-triplet oscillations possible for isotropic hyperfine coupling as the sub-spaces of different symmetry remain unconnected and as consequence, $X_x^I$ and $X_z^I$ remains zero. As the  anisotropy is included in the Hamiltonian, a non-zero signal is generated along with appearance of spike-features with the most prominent around $\theta = 90^\circ$. Again, the amplitude of this spike is  within the sensitivity achieved for $< 10$nm deep NV centres. Now we analyze the behaviour of the generated signal and spike as a function of various parameters of the RP Hamiltonian. 

The amplitude of the spike increases and it becomes narrower as the principal components $A_{xx}$ and $A_{yy}$ are increased, similar to the behaviour observed in the singlet yield in Ref. \cite{hiscock2016quantum}. This observation hints that the origin of the spike might also be similar, i.e, an avoided crossing of energy levels as function of direction of the magnetic field. However, further investigations are required to draw firm conclusions.

\begin{figure}
	\begin{center}
		\includegraphics[trim=3.2cm 2.1cm 2.7cm 2.8cm,clip,width=8.2cm]{{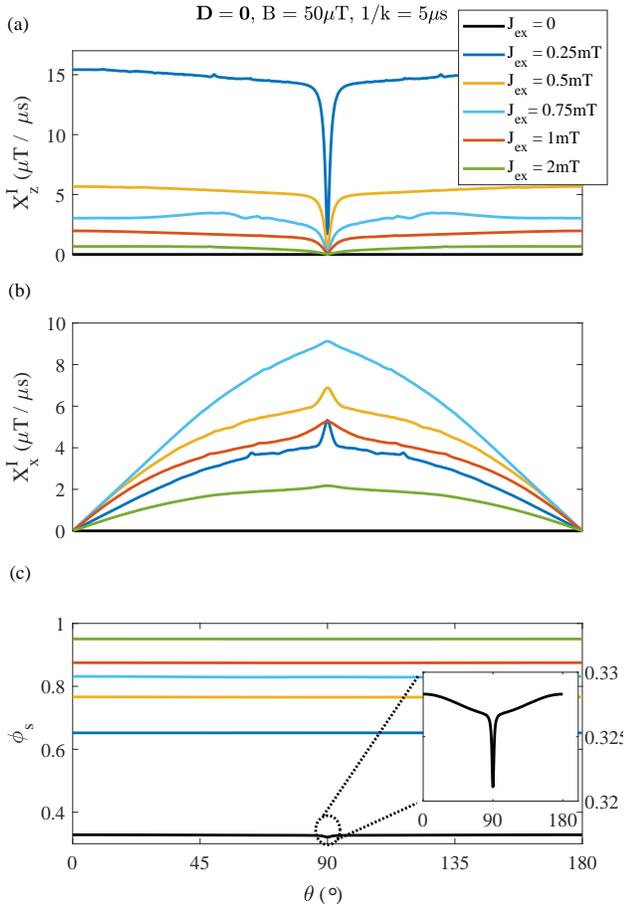}}
	\end{center}
	\caption{\small The generated signal $X_x^I$ and $X_z^I$,  and singlet yield \cite{timmel1998effects}   for various values of exchange coupling for the model system. Here $\bm{A_1} = \bm{A_2}$ and chosen according to the case `axial$^3$' (see caption of Fig. \ref{hyp_vary}). }
	\label{J_vary}
\end{figure}

In Fig  \ref{J_vary}, we show the generated signal $X_x^I$ and $X_z^I$ along with the singlet yield $\phi_s = k dt \sum_t \mathrm{Tr} [\rho_{RP} (t) |S_0\rangle\langle S_0|]$  \cite{timmel1998effects} (c) with increasing value of $J_{\mathrm{ex}}$. Using the symmetry argument, as $J_{\mathrm{ex}}$ is increased, the Hamiltonian eigenstate get closer to singlet and triplet states, and reduction (increase) in the signal (singlet yield) is observed.  The spike feature in the generated signal at $\theta = 90^\circ$ is most prominent when $J_{\mathrm{ex}}$ is comparable to the $A_{xx}$ and $A_{yy}$ principal component of the hyperfine coupling. This behaviour hints that the spike in the generated signal depends on exchange interaction (and also dipolar in general) as well as anisotropic hyperfine interaction. This is in contrast to the singlet yield where the spike depends strongly on hyperfine anisotropy as it disappears even for small  $J_{\mathrm{ex}}$,  comparable to $A_{xx}$ and $A_{yy}$ parameters \cite{hiscock2017disruption}.

\begin{figure}
	\begin{center}
		\includegraphics[trim=3.3cm 2.0cm 3.1cm 3.4cm,clip,width=8.2cm]{{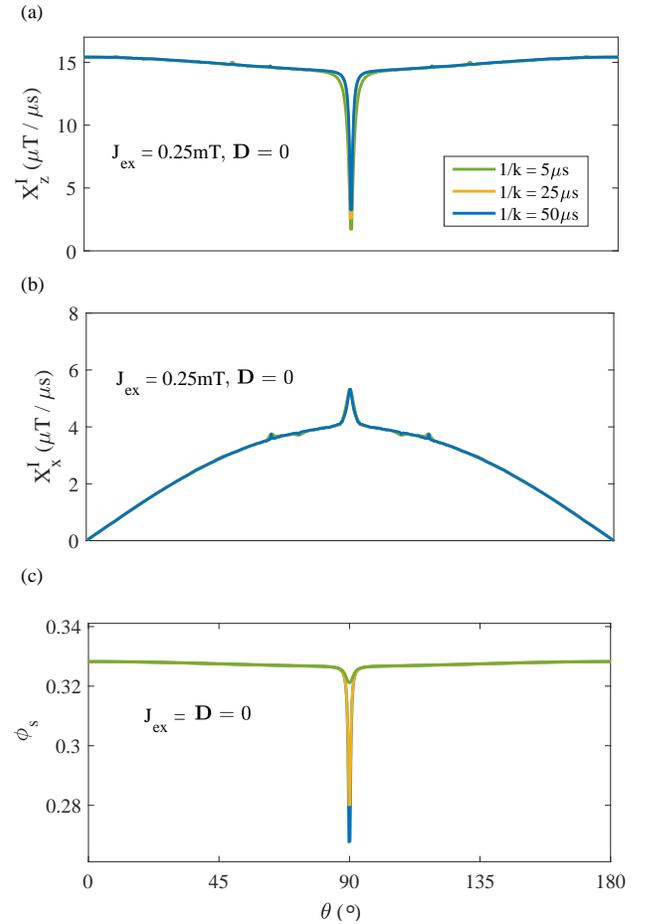}}
	\end{center}
	\caption{\small The generated signal $X_x^I$ and $X_z^I$, and singlet yield for increasing values of RP lifetime. Here $B_0 = 50\mu$T and $\bm{A_1} = \bm{A_2} $ and chosen according to the case `axial$^3$' (see caption of Fig. \ref{hyp_vary}).   }
	\label{tau_vary}
\end{figure}

In Fig  \ref{tau_vary}, we show the generated signal   $X_x^I$ and $X_z^I$ along with the singlet yield \cite{timmel1998effects} (c) with increasing value of RP lifetime $\tau$. A contrasting behaviour is observed in the generated signal and singlet yield in addition to the former being relatively less sensitive to the lifetime of the radical. This observation seems to suggest that unpaired electrons prefer to stay in the singlet state compared to other populations and coherences of the RP.  However, again, further studies are required in this direction to draw concrete conclusions which are beyond purview of current work. 

\bibliographystyle{unsrt} 
\bibliography{NV_RP}

\end{document}